# Spatially Tracking Wave Events in Partitioned Numerical Wave Model Outputs

Haoyu Jiang[1,2,3*]

[1] College of Marine Science and Technology, China University of Geosciences, Wuhan, China

[2] Laboratory for Regional Oceanography and Numerical Modeling, Qingdao National Laboratory for Marine Science and Technology, Qingdao, China

[3] Shenzhen Research Institute, China University of Geosciences, Shenzhen, China

Corresponding Author: Haoyu Jiang (Haoyujiang@cug.edu.cn)

# ABSTRACT

Numerical wave models can output partitioned wave parameters at each grid point using a spectral partitioning technique. Because these wave partitions are usually organized according to the magnitude of their wave energy without considering the coherence of wave parameters in space, it can be difficult to observe the spatial distributions of wave field features from these outputs. In this study, an approach for spatially tracking coherent wave events (which means a cluster of partitions originated from the same meteorological event) from partitioned numerical wave model outputs is presented to solve this problem. First, an efficient traverse algorithm applicable for different types of grids, termed breadth-first search, is employed to track wave events using the continuity of wave parameters. Second, to reduce the impact of the garden sprinkler effect on tracking, tracked wave events are merged if their boundary outlines and wave parameters on these boundaries are both in good agreement. Partitioned wave information from the Integrated Ocean Waves for Geophysical and other Applications dataset is used to test the performance of this spatial tracking approach. The test results indicate that this approach is able to capture the primary features of partitioned wave fields, demonstrating its potential for wave data analysis, model verification, and data assimilation.

# 1. Introduction

The ocean wave field at a given point is usually the superposition of a local windsea system and one or more swell systems originating from different meteorological events. Each of these wave systems can be considered to be independent of the others. Efforts are made to improve spectral partitioning schemes that isolate the information of different wave systems in directional wave spectra (e.g., *Gerling*, 1992; *Hanson and Phillips*, 2001; *Portilla et al.*, 2009; *Ailliot et al.*, 2013), because wave parameters integrated over the whole spectrum (such as the overall significant wave height (SWH), mean wave period, and mean wave direction) can be misleading in mixed seas. After years of development, the spectral partitioning scheme is used in increasing numbers of applications (e.g., *Collard et al.*, 2009; *Delpey et al.*, 2010; *Jiang et al.*, 2016, 2017). Some numerical wave models (NWMs) use spectral partitioning to output partitioned wave parameters, such as partitioned SWH (PSWH), partitioned peak wave direction (PPWD), and partitioned peak wave period (PPWP). For example, the WAVEWATCH III® (WW3) uses the method of *Hanson and Phillips* (2001) to partition wave spectra into at most one windsea system and five swell systems (*Tracy et al.* 2007; *The WAVEWATCH III® Development Group*, 2016). A similar data product is also available from the European Centre for Medium Range Weather Forecast ocean wave model with at most one windsea system and three swell systems (*Bidlot*, 2016).

The windsea partition in NWM output is usually labeled as Partition 0, and the swell partitions are usually sorted and labeled according to their respective SWHs. Therefore, the coherence of the wave parameters at two points close to each in space and time cannot be guaranteed for the same partition label. For instance, Figure 1 shows the global distributions of PSWH, PPWD, and PPWP for the windsea partition and the first five most energetic swell partitions on 1-January-2016, 0000 UTC from a WW3 output field (the data used here will be introduced in section 3). Discontinuity can be observed in all partitions for all three parameters, especially in partitions with larger label numbers, making it difficult to observe the features of wave fields in most of these plots. Meanwhile, the continuity of these parameters might skip from one partition to another, even between a windsea partition and a swell partition. This figure is similar to Figure 2 in *Delpey et al.* (2010) where a similar problem is discussed. The time series for any given partition at any given location may also exhibit discontinuity within the same partition label (not shown here). To generate a coherent wave event, some spatial and temporal tracking algorithms are developed to track waves that originate from the same meteorological event. There are two types of wave event tracking methods. One involves back-tracking the wave based on the linear dispersion relation to test whether a group of wave parameters corresponds to a unique (e.g., *Aarnes and Krogstad*, 2001) or known (e.g., *Delpey et al.*, 2010) source. This type of method works only for "old" swells which have propagated into "far fields" while it cannot apply to windseas and young swells because the non-linear wave-wave interaction is not usually taken into consideration in backing-tracking. The other type of tracking algorithm makes use of wave parameter continuity in space-time neighboring grid points for the same wave event (e.g., *Voorrips et al.*, 1997; *Hanson and Phillips*, 2001; *Devaliere et al.*, 2009). The temporal tracking scheme proposed by *Hanson and Phillips* (2001) is widely used to track time series of wave events (e.g., *Jiang et al.*, 2016), and a spatial tracking scheme is proposed by *Devaliere et al.* (2009) using a "spiral searching" algorithm.

Both the temporal and spatial tracking schemes are implemented in WW3 (*Van der Westhuysen et al.*, 2013; *The WAVEWATCH III® Development Group*, 2016).

Temporal tracking is conducted on a one-dimensional time series so that the search strategy can be simply searching from one time point to the next, but spatial tracking of wave events is conducted on a two-dimensional "map" so that different search algorithms can be applied. *Devaliere et al.* (2009) develop a "spiral search" algorithm to merge wave components into wave events. However, this algorithm is designed for rectangular grids and is not directly applicable to unstructured grids (*Devaliere et al.*, 2009; *Van der Westhuysen et al.*, 2013). In addition, the "spiral search" algorithm is often influenced by the land effect, which reduces the efficiency. The aim of this study is to present an algorithm for spatially tracking wave events that is applicable for both structured and unstructured grids.

## 2. Breadth-first tracking of wave field

The process of wave event spatial tracking involves searching for the wave components in all neighboring grid points with a small variation of wave parameter, which can be regarded as a process of graph traversing. A widely used traversing algorithm, breadth-first search (BFS), should therefore be a good tool for this task. BFS starts at any grid point in the wave field and explores all the neighboring grid points prior to moving on to one of them. Based on this algorithm, wave events are spatially tracked as shown in the flowchart in Figure 2 and described below:

(1) Start the search from any partition at any grid point. Put this partition into a queue (a first-in-first-out array), remove it from the original dataset, and mark this grid point as "visited". Here, the starting grid point and partition are specified as the grid point and partition with the global maximum PSWH.

(2) Designate the first element in the queue as the current wave partition. For all the partitions in all neighboring grid points that are not "visited", calculate the spectral distance gradients between them and the current partition. The spectral distance gradient is defined as:

$$\delta = \frac{1}{d}\sqrt{\left(\frac{H_{cur}-H_{nei}}{H_a}\right)^2 + \left(\frac{T_{cur}-T_{nei}}{T_b}\right)^2 + \left(\frac{|\theta_{cur}-\theta_{nei}|-\theta_{turn}}{\theta_c}\right)^2} \qquad (1)$$

where $H$, $T$, and $\theta$ are PSWH, PPWP, and PPWD, respectively, and the subscript *cur* and *nei* denote the current and neighboring points, respectively. $d$ is the geographical distance between the two points, and $H_a$, $T_b$, and $\theta_c$ are weighting factors that need to be tuned. $\theta_{turn}$ is also a tuning parameter introduced here because the nominal direction of a wave packet can change while propagating along a great circle especially at high latitudes.

(3) For each neighboring grid point, if the spectral distance gradient of the partition with minimum $\delta$ is less than an assigned threshold $\delta_{th}$, put this partition into the aforementioned queue, mark this grid point as "visited", and remove this partition from the original dataset. The "neighboring grid point" does not have to be connected with the current grid point: it can also be referred to as the grid points within a certain distance of the current point (e.g., less than 100 km). In this study, the grid points in the 8-connectivity of the current point are selected as the "neighboring grid points" for the rectangular grid, and grid points connected with the current grid point are selected for the triangular grid.

(4) When no partition meets the requirement of $\delta$ in any neighboring grid points of the current partition, the current wave partition is transferred to another array that stores the tracking results (the current wave partition is also removed from the queue).

(5) Repeat steps (2)-(4) until all the wave partitions in the queue are transferred to the array storing the results. This result array then corresponds to a tracked wave event.

(6) Mark all grid points as "unvisited", and then repeat from step (1) to track other wave events until all the wave partitions are removed from the original dataset.

According to Equation 1, a small spectral distance gradient means that the PSWH, PPWP, and PPWD are geographically changing slowly. It is noted that there is no consensus on the definition of the spectral distance in Equation 1 (e.g., *Hanson and Phillips*, 2001, *Delpey et al.*, 2009; *Husson*, 2012). Here, a form similar to the one given by *Hanson and Phillips* (2001) is selected, but other definitions should also work. The values of the tuning parameters, $H_a$, $T_b$, $\theta_c$, and $\theta_{turn}$, as well as the assigned threshold $\delta_{th}$, are shown in Table 1. During the tuning process, the value of $T_b$ is set to 1 s, and the other parameters are obtained through an iterative trial process. These values are with only one significant digit because changing these values slightly have a very small impact on the tracking results. The value of $\theta_{turn}$ is set to be proportional to the square of latitude because the nominal direction change along a great circle is larger at high latitudes. The value of $\delta_{th}$ is also regarded as being proportional to $H_{cur}$ because newly generated waves usually have both larger wave heights and larger gradient of wave parameters.

## 3. Test and improvement

Since the search process needs to compare only the wave parameters in neighboring grid points, the tracking method presented in the above section should be applicable to any partitioned NWM output in any type of grid, no matter whether it is structured or unstructured. Because WW3 can provide standard output for partitioned wave information in NetCDF format with one windsea system and up to five swell systems, the Integrated Ocean Waves for Geophysical and other Applications (IOWAGA) dataset (*Rascle and Ardhuin*, 2013), which is a hindcast dataset of WW3, is employed in this study to test the algorithm. IOWAGA is an open-access dataset which contains the global information of partitioned SWH, PWP, and PWD with a relatively high spatial-temporal resolution of $0.5°\times0.5°\times3h$. This dataset is selected here for two primary reasons: (1) The data is available for free from the FTP server of IFREMER where detailed information can also be found. (2) This dataset shows good agreement with observations from buoys and altimeters (e.g., *Ardhuin et al.*, 2010) and is selected in some other studies of swell event tracking (e.g., *Delpey et al.*, 2010; *Jiang et al.*, 2016). Figure 1 displays the information contained in this dataset, and these data are also employed for the demonstration of the tracking results. The dataset is organized in a structured grid. To test the performance of the tracking method in an unstructured grid, the data is interpolated into a global triangular grid generated by Gridgen software using the nearest neighbor approach. Other interpolation approaches, such as linear or Kriging interpolation, are not used here due to the discontinuity of wave parameters in the same partition label, or to say, interpolation approaches involving more than one grid point have to be implemented after the spatial tracking of wave events. There is almost no difference between the results for the structured and unstructured grids.

Three examples of the spatial tracking results are shown in Figure 3. The first example is a storm-generated wave event in the North Pacific. A winter storm generates waves with a PSWH of more than 6 m in the center of the wave field. The largest PPWP (~15s) is observed in two locations: one corresponds to the largest PSWH, which is primarily due to the strong wave-wave interaction in windseas, and the other is near the spatial boundary of the wave event along the wave direction, which is primarily due to the frequency dispersion of swells. Parts of this event should be regarded as windseas while parts of it should be regarded as swells, and this event implies that there is no clear distinction between the two types of wind-waves. The second example is a pure swell case which can be identified from the rainbow-like pattern of PPWP: a clear PPWP gradient is observed along the propagation direction of the wave field but the gradient is almost zero along the cross-propagation direction. When waves propagate far away from their generating areas, the spectra become narrow and the wave-wave interactions are negligible. The swells with higher PPWPs travel faster than those with lower PPWPs because of the dispersion relation, resulting in the rainbow-like period striping. The SWH distribution in the second case shows clear anisotropy which is discussed in *Delpey et al.* (2010), and the highest wave energy is found in the intermediate periods of the wave event. The third example is a typical windsea event (as can be seen in Figure 1a and 1b) generated by the winter monsoon in the South China Sea. Since the fetch is short in such a semi-enclosed sea, it is difficult for the frequency dispersion to separate the energy with different periods or for the wave to reach a fully-developed state. Therefore, the distribution of PPWP is relatively homogeneous in this event and only varies in the range of 9~10 s.

Using this method, tens of such wave events are spatially tracked at the same time. During the test, however, it is found that the same swell event might be wrongly separated into several fields with different PPWPs and PSWHs due to the garden sprinkler effect (GSE) of frequencies along swell propagation directions (*Tolman*, 2002). An example is shown in Figure 4 where the spatial boundaries of "three" wave events seem to be in good agreement with each other. The PPWDs of them are generally the same, and the PSWHs/PPWPs of them decrease/increase gradually along the wave propagation direction. Clearly, they belong to the same swell event. Although the GSE can be greatly alleviated by some numerical schemes (e.g., *Tolman*, 2002), it is still well-defined in the model output especially for swells in the far field where the spectra are very narrow. In this case, the GSE leads to a PPWP difference of ~0.5 s between adjacent grids. This difference corresponds to a spectral distance gradient of $10^{-2}$ order of magnitude, which is much larger than the $\delta_{th}$. Simply increasing the value of $\delta_{th}$ can partially solve the problem, but it will also lead to other problems as shown in Figure 5 where the $\delta_{th}$ is set to $5\times10^{-3} + 5\times10^{-3}H_{cur}$ km$^{-1}$, five times the original value. For visual analysis, results like Figure 5 are acceptable as the wave parameters in Figure 5 are already much more organized and continuous compared to Figure 1. However, distinct discontinuity of wave parameters is observed especially in the PPWP. For instance, the forerunners of some swell events come into contact with the lower frequency components of some other swell events. Also, it is noted that the three wave events in Figure 3 are regarded as the same event in Figure 5 when the $\delta_{th}$ is enlarged.

To better separate different wave events while minimizing the impact of the GSE, a corresponding procedure termed "event merging" is applied for the wave events of which the boundaries are in good agreement. To determine whether two tracked events can be merged, the following criteria are used:

(1) *Overlap Test*: If the spatial extents of the two wave events have significant overlap, they will not be regarded as being from the same wave event. In this case, the threshold grid number of overlap is set to 1‰ of the grid number of the wave event with the lower grid number.

(2) *Boundary Test*: Extend the spatial extent of one of the two events for two grid points (i.e., dilate the boundary of one of the events by two grid points from a graphics point of view). If the extended spatial extent does not have significant overlap with the spatial extent of the other event (i.e., the boundaries of two tracked wave events are inconsistent), the two events cannot be merged. Here, this threshold grid number of overlap is simply set to 10 which works well for the test dataset.

(3) *Wave Parameter Test*: If the spatial distance between the two points respectively from two to-be-merged tracked wave events which are not filtered by the above two criteria is within 150 km, they are referred to as a point pair. If the mean spectral distance gradient of all point pairs is within a threshold $\delta_{th2}$, the two tracked wave events will be merged. Here, the value of $\delta_{th2}$ is set to $4\times10^{-3}+2\times10^{-4}\ N_{ov}$ (in $km^{-1}$) where $N_{ov}$ is the number of overlapping grid points in the above *Boundary Test* (2). The idea of introducing $N_{ov}$ is that the possibility of the two tracked wave results being from the same events will be larger if their boundaries are in better agreement. This value works well for this dataset, and changing it within ±20% has almost no influence on the results due to the first two criteria.

After merging, the three tracked results in Figure 4 are merged into one in Figures 6a and 6b, showing that they are indeed the same event. Figures 6c and 6d show the same event of the one in Figures 3c and 3d, but the spatial extent of the event appears to be larger after merging the long-PPWP stripes, and some data gaps are filled (e.g., the region with the maximum PSWH), showing the improvement of the tracking results using this merging procedure. Figures 6e and 6f are also a case in which several tracked results are merged into one wave event. In this case, a seam can be observed in the low-frequency (long-period) part of the wave event. That is the reason why the spatial extent of one of the tracked events is extended for two grid points instead of one in the above *Boundary Test* (2). The PSWH near this seam is low (less than 0.2 m) so that the GSE might lead to lower wave energy in this seam which might not be identified by the partitioning algorithm of *Hanson and Phillips* (2001). These results all show that the GSE in the frequency direction has a relatively large impact on swells with long periods in the far field. Although this effect only has a very small impact on the simulation of overall SWHs due to the low energy contribution of such partitions, it should be taken into account in tracking and modeling the forerunners of ocean swells.

## 4. Summary and Discussion

Wave partitions in NWM output are usually organized without considering the coherence of wave parameters in space, making it difficult to observe the features of wave fields. In this study, an approach for spatially tracking coherent wave events using partitioned NWM output is presented to solve this problem. This two-step method first traverses the wave event using the continuity of wave parameters by BFS, then merges the events of which the boundaries and the wave parameters on the boundaries are both in good agreement. The BFS in the first step is an efficient traversal method which works for all grid types. The second step is to reduce the impact of the GSE, which effectively improves the tracking results. To have a better global view of the tested wave field, the first 100 largest (with respect to area) tracked wave events on 1-January-2016, 0000 UTC are shown in Figure 7. The wave events without overlap with respect to the dilated spatial extents are plotted on the same map. Although more than six maps are used to display the information of the wave field (Figure 1 only uses six maps), the wave field features are much more clearly-observed and more coherent in Figure 7 than those in Figure 1.

Although the results seem to be nice for some events, there are still two shortcomings in this approach that the author fails to overcome at this stage. On the one hand, not all the wave partitions are assigned to a wave event and the spatial extents of some tracked events are small (only a few grid points, as shown in Figure 7). This is largely due to the error of spectral partitioning algorithm itself. It is noted that the WW3 retains at most six partitions at a given location, but sometimes more than six partitions can coexist. In addition, the watershed-algorithm-based partitioning scheme (e.g., *Hanson and Phillips*, 2001) is a purely morphological method that does not consider wave dynamics. When the energy peak of a partition becomes smaller than the tail of a nearby partition, it might be overwhelmed by the nearby partition, and the watershed-algorithm-based partitioning scheme cannot identify it (*Ailliot et al.*, 2013). Both of these might lead to discontinuity of partitioned wave parameters in neighboring grid points and interrupt the procedure of wave event tracking. On the other hand, only a single partition can be tracked at a given location, because only the partition with the minimum spectral distance gradient at that point can be included in the wave event according to this tracking method. However, this is not always the case because the complicated space-time structure of the wind field might sometimes lead to more than one partitions at the same location for the same event. Removing the step of marking grid points as "visited" and including all the partitions with $\delta$ less than $\delta_{th}$, rather than only including the partition with minimum $\delta$, can partially solve this problem.

Nevertheless, the author has not yet determined a way to visualize the results for the condition that more than one partition is in the same grid point for the same wave event.

Despite these shortcomings, this spatial tracking method presented here can capture the main features of partitioned wave fields from NWM output, and there are several potential applications for it. As the visualization of the original NWM partitioned output directly can only provide very limited information on the wave field due to the spatial incoherence of identified partitions, the tracked results for wave events make it convenient to have a quick view of the model output. The resulting spatially coherent wave fields can be used to track the spatial-temporal evolution of wave events from windseas to swells, which can be employed in the study of the space-time structure of wave fields (e.g., *Delpey et al.*, 2010). By comparing spatially-tracked model output with wave spectra measurements from buoys, this approach can also be used to identify potential error sources in wave modeling. Furthermore, this approach might serve as a potential tool for the data assimilation of partitioned NWM output: if coherent evolution of wave events in the NWMs can be obtained, wave data assimilation can be conducted at the "event" level by assimilating in-situ measurements from buoys or remote sensing data from synthetic aperture radars into the corresponding wave events. With the increasing attention being paid to partitioned wave information in NWMs, such a spatial tracking approach could contribute to both the scientific community and the marine industries.

**Acknowledgments**

# Figure & Table Captions

**Table 1.** The values of the tuning parameters.

**FIG. 1.** Modeled global distributions of PSWH (left), PPWP (right), and PPWD (arrows) for the windsea (1$^{st}$ row) partition and the first (2$^{nd}$ row) to the fifth (6$^{th}$ row) swell partitions on 1-January-2016, 0000 UTC. The blank areas mean no corresponding partition is detected. The arrows in all figures are not scaled with the background field.

**FIG. 2.** The flowchart for spatial tracking of wave events. Each color represents a step described in section 2b with the legend shown in the upper-left (S denotes STEP).

**FIG. 3.** Distributions of PSWH (upper), PPWP (lower), and PPWD (arrows) for three examples of the wave events tracked by the method described in Section 2 on 1-January-2016, 0000 UTC. Subplots a and b (left) show a storm-generated wave event in the North Pacific. Subplots c and d (middle) show a swell event which has propagated out of the generating areas. Subplots e and f (right) show a windsea event in the South China Sea.

**FIG. 4.** An illustration of the distributions of PSWH (upper), PPWP (lower), and PPWD (arrows) for a swell event which is incorrectly separated by the tracking method described in Section 2 on 1-January-2016, 0000 UTC.

**FIG. 5.** Distributions of PSWH (left), PPWP (right), and PPWD (arrows) for a wave event when the $\delta_{th}$ is set to five times the original value in Table 1.

**FIG. 6.** Distributions of PSWH (upper), PPWP (lower), and PPWD (arrows) for three examples of wave events on 1-January-2016 after event merging using the criteria in Section 3. Subplots a and b (left) are the same event as in Figure 4. Subplots c and d (middle) are the same event as in subplots c and d of Figure 3. Subplots e and f (right) show another merged swell event where a seam can be observed in the region with long PPWPs.

**FIG. 7.** Distributions of PSWH (left), PPWP (right), and PPWD (arrows) for the first 100 largest (with respect to area) tracked wave events on 1-January-2016, 0000 UTC. The wave events without overlap with respect to dilated spatial extents are plotted on the same map.

**Table 1.** The values of the tuning parameters.

| Parameters | Values (Unit) |
|---|---|
| $H_a$ | 2 (m) |
| $T_b$ | 1 (s) |
| $\theta_c$ | 10 (°) |
| $\theta_{turn}$ | Max $[5 \times \sin^2(\varphi), |\theta_{cur} - \theta_{nei}|]$ (°) * |
| $\delta_{th}$ | $10^{-3} + 10^{-3} H_{cur}$ (km$^{-1}$) |

\* $\varphi$ is the latitude of the current grid point.

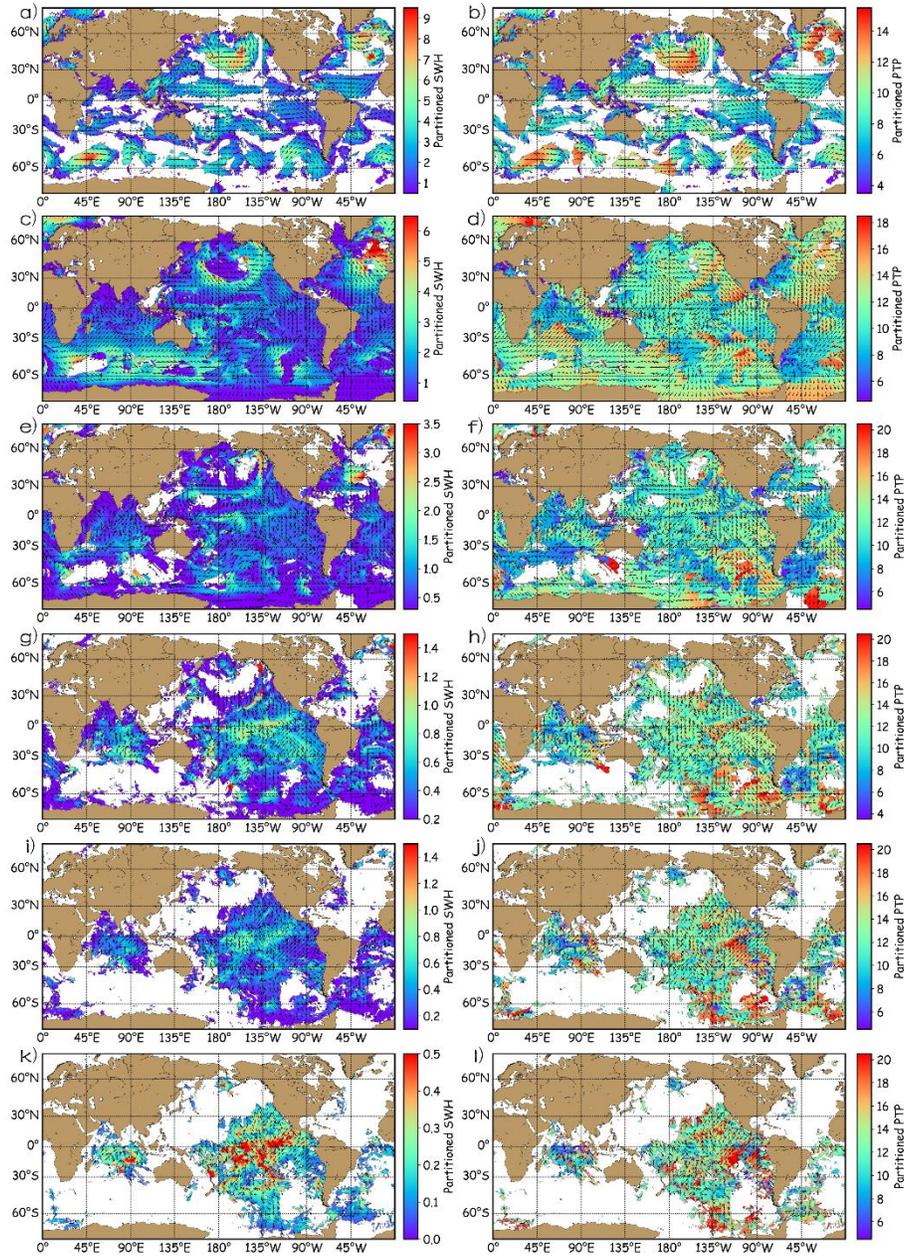

**FIG. 1.** Modeled global distributions of PSWH (left), PPWP (right), and PPWD (arrows) for the windsea (1st row) partition and the first (2nd row) to the fifth (6th row) swell partitions on 1-January-2016, 0000 UTC. The blank areas mean no corresponding partition is detected. The arrows in all figures are not scaled with the background field.

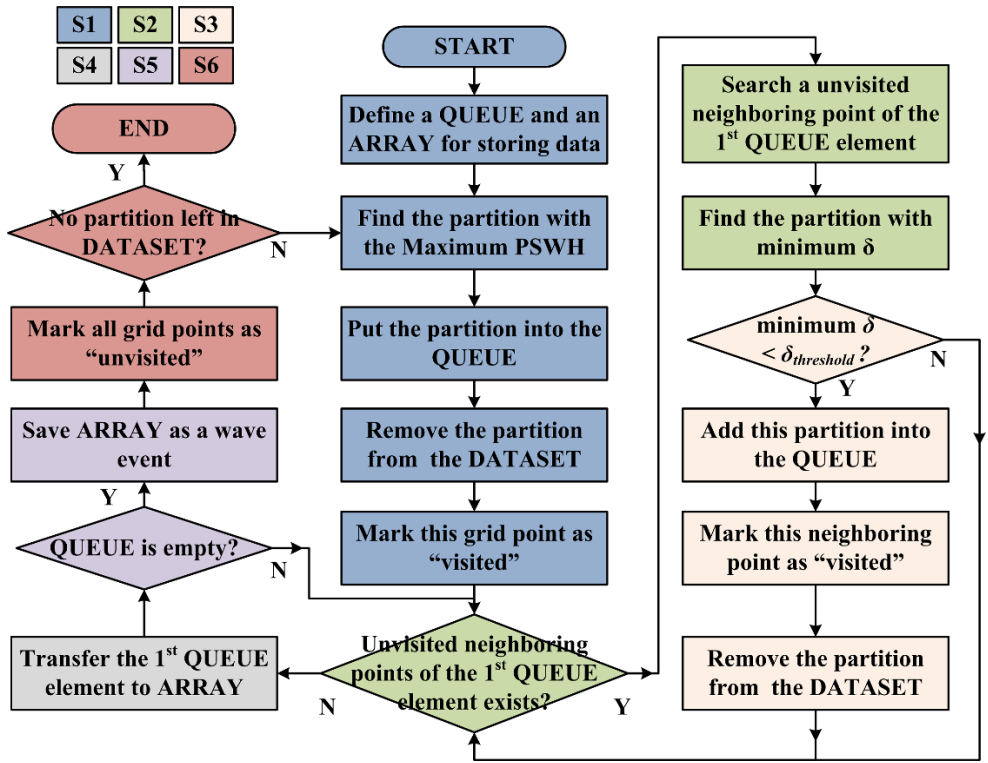

**FIG. 2.** The flowchart for spatial tracking of wave events. Each color represents a step described in section 2b with the legend shown in the upper-left (S denotes STEP).

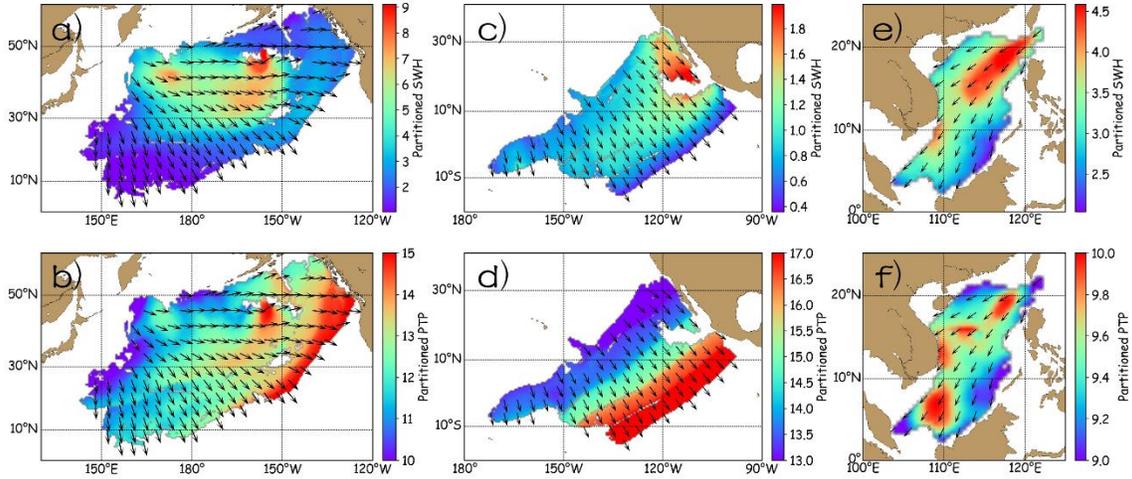

**FIG. 3.** Distributions of PSWH (upper), PPWP (lower), and PPWD (arrows) for three examples of the wave events tracked by the method described in Section 2 on 1-January-2016, 0000 UTC. Subplots a and b (left) show a storm-generated wave event in the North Pacific. Subplots c and d (middle) show a swell event which has propagated out of the generating areas. Subplots e and f (right) show a windsea event in the South China Sea.

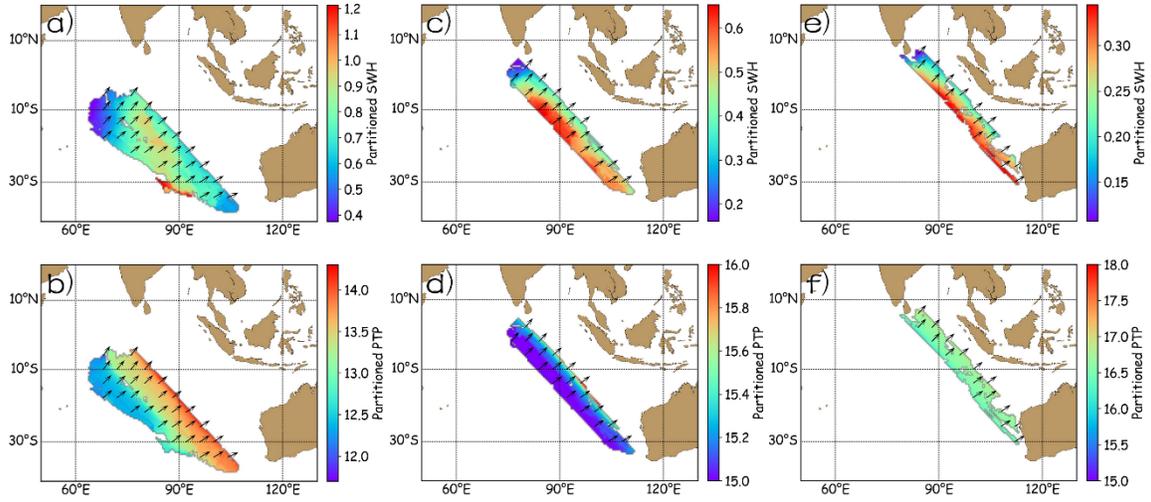

**FIG. 4.** An illustration of the distributions of PSWH (upper), PPWP (lower), and PPWD (arrows) for a swell event which is incorrectly separated by the tracking method described in Section 2 on 1-January-2016, 0000 UTC.

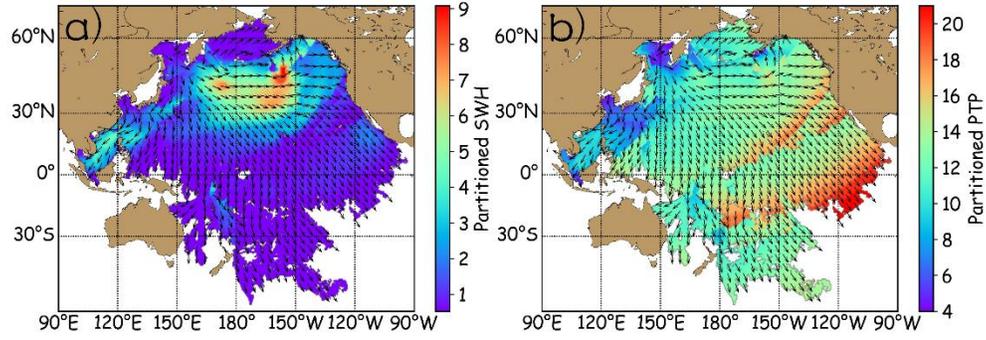

**FIG. 5.** Distributions of PSWH (left), PPWP (right), and PPWD (arrows) for a wave event when the $\delta_{th}$ is set to five times the original value in Table 1.

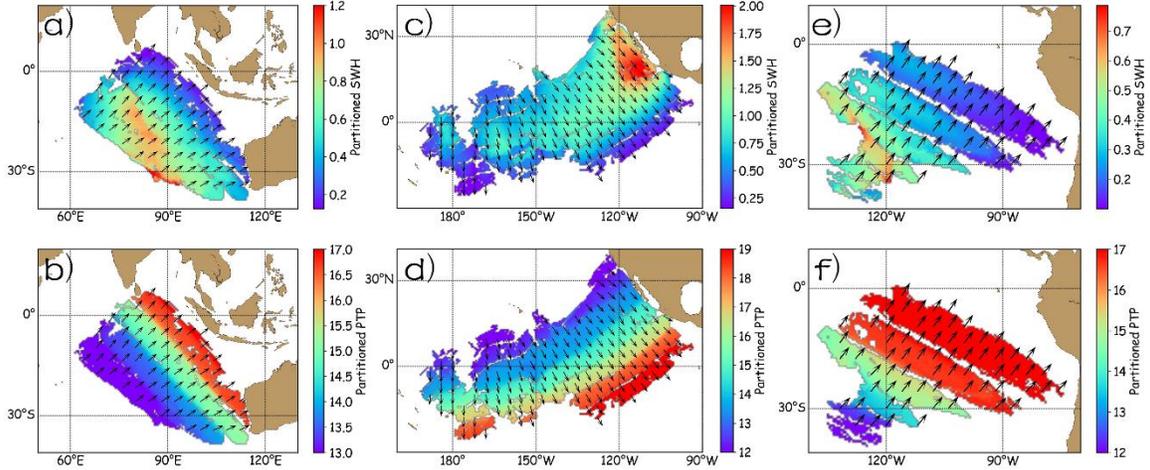

**FIG. 6.** Distributions of PSWH (upper), PPWP (lower), and PPWD (arrows) for three examples of wave events on 1-January-2016 after event merging using the criteria in Section 3. Subplots a and b (left) are the same event as in Figure 4. Subplots c and d (middle) are the same event as in subplots c and d of Figure 3. Subplots e and f (right) show another merged swell event where a seam can be observed in the region with long PPWPs.

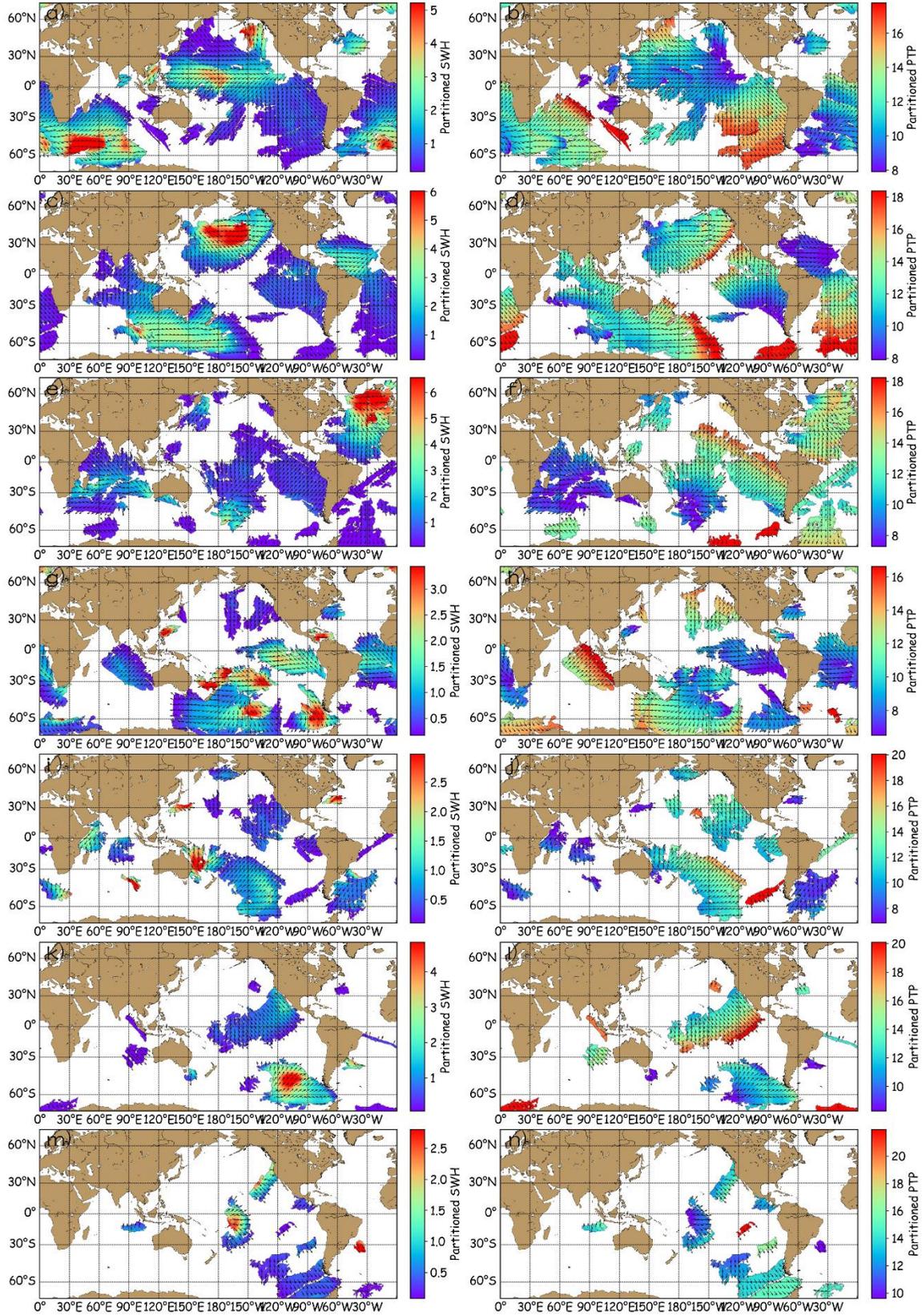

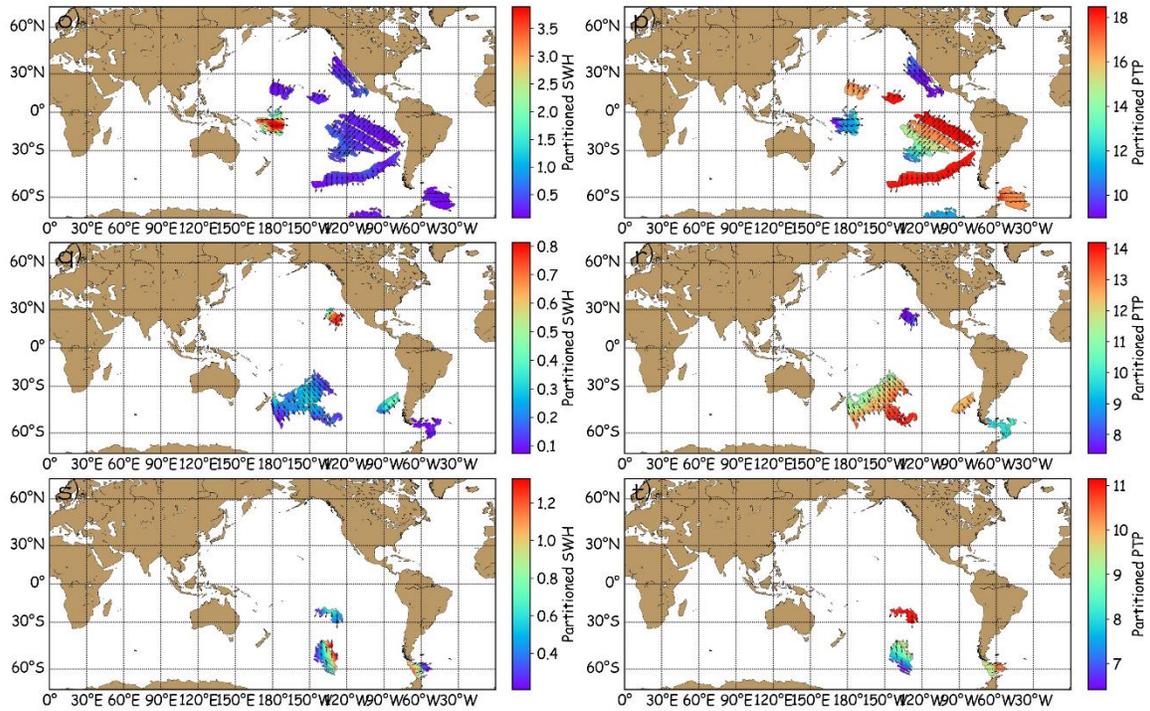

**FIG. 7.** Distributions of PSWH (left), PPWP (right), and PPWD (arrows) for the first 100 largest (with respect to area) tracked wave events on 1-January-2016, 0000 UTC. The wave events without overlap with respect to dilated spatial extents are plotted on the same map.